\documentclass[
twocolumn
]{ceurart}

\usepackage{listings}
\usepackage{ulem}
\usepackage{arydshln}
\usepackage{xcolor}
\usepackage{float}
\usepackage{amsmath}
\usepackage{multirow}
\usepackage{longtable}

\begin{document}

\copyrightyear{2022}
\copyrightclause{Copyright for this paper by its authors.
  Use permitted under Creative Commons License Attribution 4.0
  International (CC BY 4.0).}

\conference{SIGIR'24 Workshop on Information Retrieval's Role in RAG Systems, July 18, 2024, Washington D.C., USA.}

\title{Multi-Aspect Reviewed-Item Retrieval via LLM Query Decomposition and Aspect Fusion}


\author[1]{Anton Korikov}[%
orcid=0009-0003-4487-9504,
email=anton.korikov@mail.utoronto.ca,
]
\cormark[1]
\fnmark[1]
\address[1]{University of Toronto, Toronto, Canada}

\author[1]{George Saad}[%
orcid=0009-0000-3549-9874,
email=g.saad@mail.utoronto.ca,
]
\fnmark[1]

\author[1]{Ethan Baron}[%
orcid=0009-0004-2461-5760,
]

\author[1]{Mustafa Khan}[%
orcid=0009-0008-3622-7270,
email=mr.khan@mail.utoronto.ca,
]

\author[1]{Manav Shah}[%
orcid=0009-0008-4728-0771
]

\author[1]{Scott Sanner}[%
orcid=0000-0001-7984-8394,
]

\cortext[1]{Corresponding author.}
\fntext[1]{These authors contributed equally.}

\begin{abstract}
While user-generated product reviews often contain large quantities of information, their utility in addressing natural language product queries has been limited, with a key challenge being the need to aggregate information from multiple low-level sources (reviews) to a higher item level during retrieval. Existing methods for reviewed-item retrieval (RIR) typically take a late fusion (LF) approach which computes query-item scores by simply averaging the top-K query-review similarity scores for an item. However, we demonstrate that for \textit{multi-aspect queries} and \textit{multi-aspect items}, LF is highly sensitive to the distribution of aspects covered by reviews in terms of aspect frequency and the degree of aspect separation across reviews. To address these LF failures, we propose several novel aspect fusion (AF) strategies which include Large Language Model (LLM) query extraction and generative reranking. Our experiments show that for imbalanced review corpora, AF can improve over LF by a MAP@10 increase from $0.36\pm0.04$ to $0.52\pm0.04$, while achieving equivalent performance for balanced review corpora.
\end{abstract}


\maketitle

\section{Introduction}
User-generated reviews are an abundant and rich source of data that has the potential to be used to improve the retrieval of reviewed-items such as products, services, or destinations. However, a challenge of using review data for retrieval is that information has to be aggregated across multiple (low-level) reviews to a (higher) item-level during retrieval. Recent work \cite{abdollah2023self}, defining this Reviewed-Item Retrieval setting as RIR, showed that state-of-the-art results could be achieved by using a bi-encoder to aggregate review information to an item-level in a process called late fusion (LF). As opposed to aggregating review information to an item-level \textit{before} query-scoring (early fusion), LF first computes query-review similarity to avoid losing information before scoring, and then averages the top-$K$ query-review similarity scores to get a query-item similarity score. Recently, LF has been implemented by retrieval augmented generation (RAG) driven conversational recommendation (ConvRec) systems for generative recommendation, explanation, and interactive question answering \cite{Kemper2024}. 

In this paper, we extend RIR to a multi-aspect retrieval setting, formulating what we call multi-aspect RIR (MA-RIR). In this problem, our goal is to retrieve relevant items for a multi-aspect query by using the reviews of multi-aspect items. Specifically, for an item with multiple aspects, we assume that each review describes at least one, and up to all, of the item's aspects. 

As our primary contributions:
\begin{itemize}
    \item We formulate the MA-RIR problem and identify failure modes of LF under imbalanced  review-aspect distributions, considering imbalances due to both aspect frequency and the degree of aspect separation across reviews.
    \item We propose several novel aspect fusion strategies, which include LLM query extraction and reranking, to address failures of LF review-score aggregation on imbalanced multi-aspect review distributions. 
    \item We leverage a recently released multi-aspect retrieval dataset, Recipe-MPR \cite{10.1145/3539618.3591880}, with ground-truth query- and item- aspect labels to generate four multi-aspect review distributions with various aspect balance properties, and numerically evaluate the effect of review-aspect balance on MA-RIR.
    \item Our simulations show that for imbalanced data, Aspect Fusion can improve over LF by MAP@10 increase from $0.36\pm0.04$ to $0.52\pm0.04$ while achieving equivalent performance for balanced data.
    \item We show that LLM reranking in both cross-encoder and zero-shot (ZS) listwise reranking settings can provide some improvements when given a large enough number of reviews, but risk decreasing performance when not enough reviews are provided.
\end{itemize}

\section{Background}
\subsection{Neural IR}
Given a set of documents $\mathcal{D}$ and a query $q \in \mathcal{Q}$, an IR task $IR\langle \mathcal{D},q \rangle$ is to assign a similarity score $S_{q,d} \in \mathbb{R}$ between the query and each document $d \in \mathcal{D}$ and return a ranked list of top scoring documents. The standard first-stage neural-IR method \cite{reimers2019sentence} for a large corpus is to first use a bi-encoder $g(\cdot): \mathcal{Q} \cup \mathcal{D} \rightarrow \mathbb{R}^m$ to map a query $q$ and document $d$ to their respective embeddings $g(q) = \mathbf{z}^q$ and $g(d) = \mathbf{z}^d$. A similarity function $f(\cdot,\cdot):\mathbb{R}^m \times \mathbb{R}^m \rightarrow \mathbb{R}$, such as the dot product, is then used to compute a query-document score $S_{q,d} = f(\mathbf{z}^q,\mathbf{z}^d)$. For web-scale corpora, exact similarity search for the top query-document scores is typically impractical, so approximate similarity search algorithms \cite{johnson2019billion} are used instead.

\subsection{Reviewed-Item Retrieval}
\subsubsection{Problem Formulation}  \label{sec:background}
Information retrieval across two-level data structures was previously studied by Zhang and Balog \cite{zhang2017design}. Specifically, Zhang and Balog define the \textit{Object Retrieval} problem, where (high-level) objects are described by multiple (low-level) documents. Given a query, the task is to retrieve high-level objects by using information in the low-level documents.

To investigate a special case of object retrieval where the goal is retrieving items (e.g., products, destinations) based on their reviews, Abdollah Pour \textit{et al.} \cite{abdollah2023self} recently proposed the Reviewed-item Retrieval (RIR) problem. In the $RIR\langle  \mathcal{I}, \mathcal{D}, q \rangle$ problem, there is a set of items $\mathcal{I}$, where every item $i$ is a high-level object. Each item is described by a set of reviews (i.e., ``low-level documents") $\mathcal{D}_i \subset \mathcal{D}$, and the $r$'th review of item $i$ is $d_{i,r} \in \mathcal{D}_i$. The main difference between RIR and Object Retrieval is that in RIR a low-level document $d_{i,r}$ cannot describe more than one high-level object $i$, while Object Retrieval allows for more general two-level structures. Given a query $q \in \mathcal{Q}$ and a score $S_{q,i}$ between $q$ and each item $i$, the goal of RIR is to retrieve a ranked list $L^q$ of top-$K_I$ scoring items:
\begin{align*}
L^q = (i_1,...,i_{K_I}) \quad  \mbox{s.t.} \; & i_1 \in \arg \max_i \{ S_{q,i}\} \\ & S_{q,i_k}, \geq S_{q,i_{k+1}}, \quad \forall i_{k} \in L^q.
\end{align*}

\subsubsection{Fusion} To get a query-item score $S_{q,i}$ using an item's review set $\mathcal{D}_i$, review information needs to be aggregated to an item level: this process is called \textit{fusion}. Two alternatives exist for fusion \cite{zhang2017design}: if low-level information is aggregated \textit{before} a query is used for scoring, it is called Early Fusion (EF) --- in contrast, if the aggregation occurs \textit{after} query-scoring, it is called Late Fusion (LF). 

For EF in RIR, Abdollah Pour \textit{et al.} \cite{abdollah2023self} experiment with mean-pooling and contrastive learning methods to create an item embedding $\mathbf{z}^i \in \mathbb{R}^m$ from review embeddings $\{ \mathbf{z}^d \}_{d \in \mathcal{D}_i}$. They then directly compute the similarity between $\mathbf{z}^i$ and a query embedding $\mathbf{z}^q$ as the query-item score $S_{q,i} = f(\mathbf{z}^q,\mathbf{z}^i)$.

For LF in RIR, these authors first compute query-review similarity scores $S_{q,d_{i,r}} = f(\mathbf{z}^q,\mathbf{z}^{d_{i,r}})$. They then aggregate these scores into a query-item score $S_{q,i}$ by averaging the top $K_R$ query-review scores for each item:
\begin{equation}\label{eqn:LF}
  S_{q,i} = \frac{1}{K_R}\sum_{r=1}^{K_R} S_{q,d_{i,r}} .
\end{equation}
 
Numerical evaluations performed for EF and LF for RIR demonstrate that EF has significantly worse performance than LF \cite{abdollah2023self}, and
Abdollah Pour \textit{et al.} conjecture that EF performs worse because it loses fine-grained review information before query-scoring. In contrast, by delaying fusion, LF preserves review-level information during query-scoring. Due to these findings, we do not study EF for MA-RIR, rather, we focus on developing Aspect Fusion as an extension of LF, discussed next.

\section{Multi-Aspect Reviewed Item Retrieval}\label{sec:data_structures}

\begin{figure}[b]
       \centering
       \includegraphics[width=0.9\linewidth]{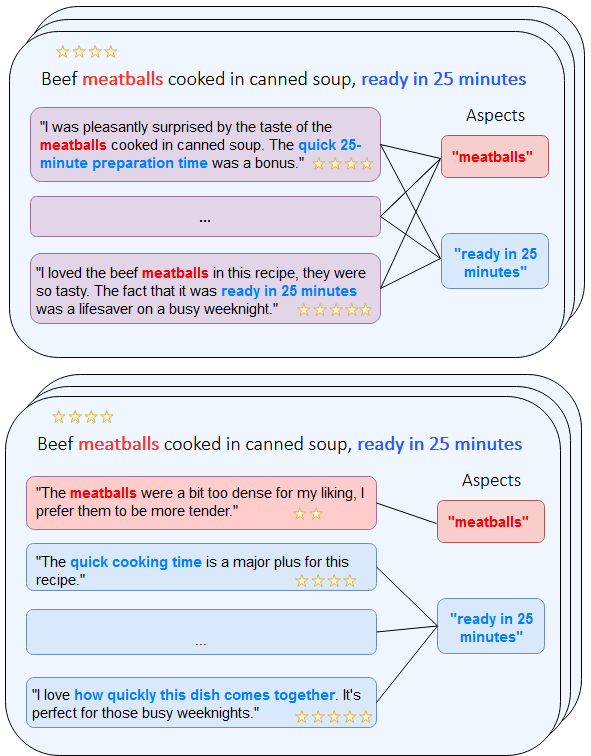}
       \caption{Two extremes of item aspect distributions, showing reviews for an item with aspects ``meatballs'' and ``ready in 25 minutes'': a) \textit{Fully overlapping }(top) --- Each review mentions all item aspects.  b) \textit{Fully disjoint with imbalanced aspect frequency }(bottom) --- no review mentions more than one aspect, and some aspects are mentioned much more frequently than others. }
       \label{fig:review_mix}
\end{figure}

\subsection{Multi-Aspect Queries}\label{ssec:multi_asp_queries}
This paper focuses on retrieving relevant items using their reviews for a multi-aspect query, such as “\textit{Can I have a \textbf{meatball} recipe that \textbf{doesn't take too long}?}”. We define a query aspect to be a sub-span of a multi-aspect query that represents a distinct topic (or facet) in the query, for instance the sub-spans ``\textit{``meatball''} and \textit{``doesn't take too long''} in the previous sentence. While there is ambiguity in identifying which sub-spans, if any, in a query should be considered aspects, this sub-span based definition is a simple way to represent aspects and is conducive to overlap-based evaluations of aspect extraction such as intersection-over union (IOU).   Formally, we denote the set of aspects in query $q$ as $\mathcal{A}^{\text{query}}_q$, where the $j'$th query aspect is $a^{\text{query}}_{q,j} \in \mathcal{A}^{\text{query}}_q$. In this work, multi-aspect queries are assumed to be logical AND queries for all aspects, though an aspect itself can represent other logical operators such as XOR (e.g. a query aspect may be ``\textit{chicken or beef}''). Finally, we assume all query aspects are equally important --- a further discussion of weighted multi-aspect retrieval can be found in Section \ref{sec:rel_work}. 
\subsection{Multi-Aspect Reviewed-Items}In addition to considering multi-aspect queries, we also consider multi-aspect items described by reviews. For instance, a multi-aspect item that is relevant to the multi-aspect query example above might be a recipe titled ``\textit{Beef \textbf{meatballs} cooked in canned soup, \textbf{ready in 25 minutes}}''.  However, since our goal is to isolate the properties of review-based retrieval, we assume that no such natural language (NL) item-level description is available. Instead, we assume that the item's aspects are described in reviews. Obviously, item-level descriptions (e.g. titles) \textit{are} often available in practice, so a prime direction for future work is fusion across multiple levels of NL data during reviewed-item retrieval.

Examples of reviews describing the item in the previous paragraph, which has aspects ``\textit{meatballs}'' and ``\textit{ready in 25 minutes}'', are shown in Figure \ref{fig:review_mix}. In this paper, we assume that a review $d_{i,r}$ must mention at least one item aspect $a^{\text{item}}_{i,j} \in \mathcal{A}^{\text{item}}_i$ and could mention up to all item aspects. Formally, the distribution of item aspects across reviews can be defined with a bipartite aspect distribution graph $\mathcal{G} = \{\mathcal{D},\mathcal{A}^{\text{item}},\mathcal{E}\}$ where an edge $(d_{i,r},a^\text{item}_{i,j}) \in \mathcal{E}$ exists if review $d_{i,r} \in \mathcal{D}_i$ mentions aspect $a^\text{item}_{i,j} \in \mathcal{A}^{\text{item}}_i$. We also let $\mathcal{A}_i^{\text{rel},q}\subseteq \mathcal{A}^{\text{item}}_i$ represent the set of item-aspects that are relevant to a query and should be considered during retrieval. We define the $MA-RIR \langle \mathcal{A} , \mathcal{E},\mathcal{D},q\rangle$ problem as the task of retrieving a ranked list of relevant multi-aspect items $L^q$ for a multi-aspect query $q$, where $\mathcal{A} = \mathcal{A}^{\text{item}} \cup \mathcal{A}^{\text{query}}$ .

\subsection{Multi-Aspect Review Distributions}
As we will demonstrate with numerical simulations on LLM-generated review data, understanding review distributions in terms of \textit{aspect frequency} and \textit{degree of aspect separation} between reviews is key to designing successful MA-RIR techniques. Figure \ref{fig:review_mix} shows two extremes of aspect distributions that are among the distributions we explore in our experiments.

\subsubsection{Fully Overlapping Distributions} Figure \ref{fig:review_mix}a) shows a \textit{fully overlapping} aspect distribution where \textit{each} review mentions \textit{all} aspects --- in this case, the bipartite graph $\mathcal{G}$ (see the RHS of Figure \ref{fig:review_mix}) is fully connected for item $i_1$. This is the most balanced review aspect distribution possible for an item, and, because of this ``perfect'' aspect balance, we postulate that aspect-agnostic retrieval approaches such as standard LF will perform competitively on such distributions. 

\subsubsection{Degree of Separation and Aspect Frequency} In contrast to the case of perfect review-aspect balance, Figure \ref{fig:review_mix}b) shows an extreme case of aspect imbalance. Firstly, one aspect is mentioned much more frequently than another --- this is an aspect frequency imbalance. Secondly, \textit{each} review mentions \textit{only one} aspect --- this is a maximal degree of separation of aspects across reviews (fully disjoint). Mathematically, $\mathcal{G}$ has $|\mathcal{A}_{i_1}^{\text{item}}|$ (disjoint) star components where some stars have a singificantly higher degree than others. In the next section, we discuss the negative effects of imbalanced review-aspect distributions on LF performance on MA-RIR, and propose aspect fusion as a method for mitigating these negative effects.



\section{Aspect Fusion for MA-RIR}\label{sec:aspect_fusion}
\begin{figure}
   \centering
\includegraphics[width=\linewidth]{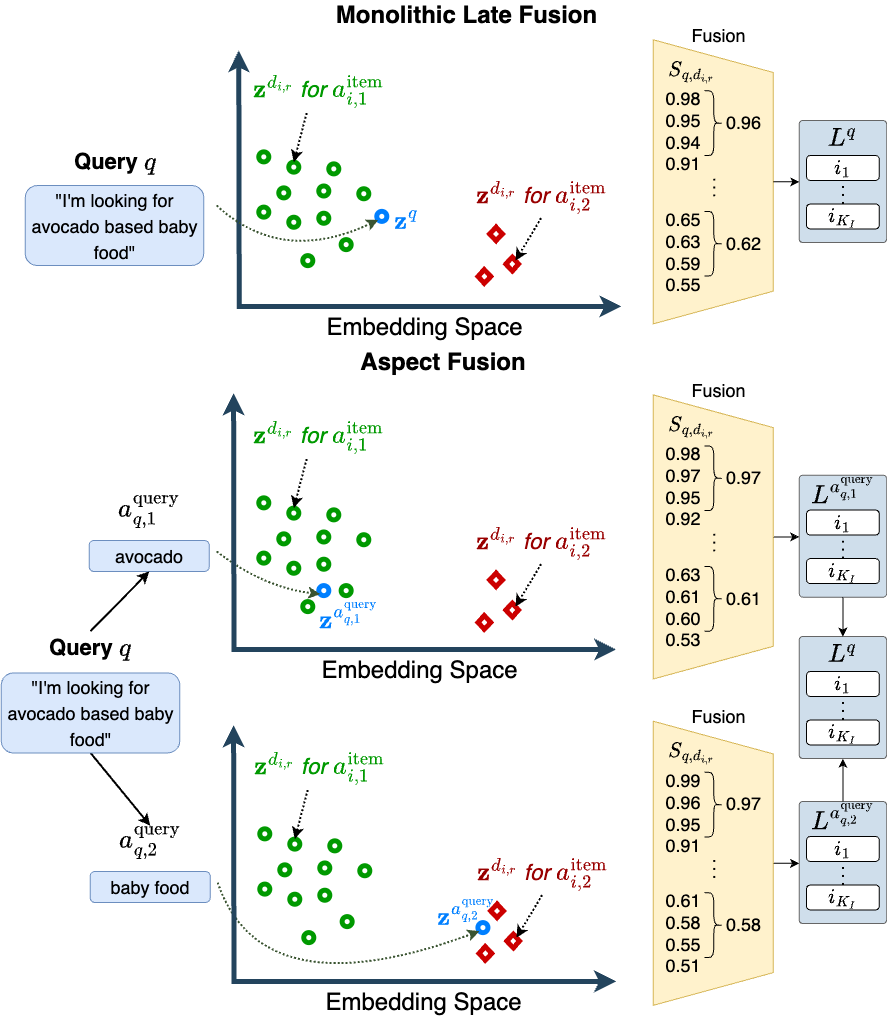}
   \caption{a) Top. In (Monolithic) LF, the full query is scored against all reviews, and the top $K_R$ query-review scores are averaged for each item to produce a query-item score. b) Bottom. Aspect Fusion extracts aspects (i.e., query subspans) from a query, performs LF with each aspect, and aggregates the resulting top $K_I$ item lists (i.e., one list per extracted aspect) to a final list.}
   \label{fig:mono_vs_aspect}
\end{figure}
\subsection{Desiderata of Aspect Fusion}
Recall that LF computes a query-item similarity score by averaging the top $K_R$ query-review similarity scores using Equation \eqref{eqn:LF}. For MA-RIR, we propose two desiderata for the aspect distribution in the top $K_R$ reviews during fusion.

\paragraph{Desideratum 1:} Since we assume multi-aspect queries are AND queries, if an item contains $\mathcal{A}_i^{\text{rel},q}$ relevant aspects for query $q$, the $K_R$ reviews used for LF should mention all $\mathcal{A}_i^{\text{rel},q}$ of those relevant aspects.

\paragraph{Desideratum 2:} As mentioned in Section \ref{ssec:multi_asp_queries}, we also assume all query aspects are equally important, which implies that aspect frequency should be identical for all $\mathcal{A}_i^{\text{rel},q}$ aspects in the top $K_R$ retrieved reviews. \\

In a fully overlapping distribution (Figure \ref{fig:review_mix}a) where \textit{each} review mentions \textit{each} aspect, both Desiderata 1 and 2 are guaranteed to be satisfied by any subset of item reviews. We thus argue that standard LF should be sufficient when reviews fully overlap in aspects, and focus on developing Aspect Fusion methods that address the failures of LF for imbalanced review-aspect distributions.

\subsection{Failures of LF under Review-Aspect Imbalance}\label{sec:lf_failures}
Standard LF will fail to achieve Desiderata 1 and 2 for review-aspect distributions with at least some degree of disjointedness and  aspect frequency imbalance under the following assumptions.  

\paragraph{Aspect Popularity Bias} 
Aspects that are reviewed more frequently are more likely to be mentioned in the top $K_R$ reviews.

\paragraph{Embedding Bias} 
The non-isotropic nature of the embedding space \cite{ethayarajh-2019-contextual} biases retrieval towards one aspect. Consider two equally sized and fully disjoint review subsets $\mathcal{D}^j_i \subset \mathcal{D}_i$ and $\mathcal{D}^k_i \subset \mathcal{D}_i$ in which reviews mention only a single aspect $a^{\text{rel}}_{i,j} \in \mathcal{A}_i^{\text{rel},q}$ or 
$a^{\text{rel}}_{i,k} \in \mathcal{A}_i^{\text{rel},q}$,  respectively, for some item $i$. If query-review similarity scores tend to be higher when a review describes aspect $a^{\text{rel}}_{i,j}$ as opposed to aspect $a^{\text{rel}}_{i,k}$, LF will be more likely to select reviews from review set $\mathcal{D}^j_i$ for the top $K_R$ fused reviews. For example, in Figure \ref{fig:review_mix}b), the reviews describing cooking time might be more likely to score higher with the full query than reviews describing ``meatballs''. 


\subsection{Aspect Fusion}
To address these failures of LF on imbalanced data, we introduce several methods for \textit{Aspect Fusion}, which explicitly utilizes the multi-aspect nature of reviews during fusion to address multi-aspect queries.

\subsubsection{Aspect Extraction}
To extract aspects from queries, we propose to use few-shot (FS) prompting with an LLM. Though the number of query-aspects is typically not known \textit{a priori}, since we study multi-aspect queries, our proposed prompt (Figure \ref{fig:aspect_extract_prompt} in the Appendix) asks that at least two non-overlapping sub-spans of the query be extracted as aspects. We represent the set of extracted query aspects for query $q$ as $\mathcal{A}_q^{\text{ext}}$ and let $A_q^{\text{e}} = |\mathcal{A}_q^{\text{ext}}|$.

\subsubsection{Aspect-Item Scoring}
The key to Aspect Fusion is directly computing \textit{aspect-review} similarity scores $S_{a,d_{i,r}}$, as opposed to similarity scores between reviews and a monolithic query, since the later can be negatively impacted by review-aspect distribution imbalance. Aspect similarity scores are computed by separately embedding each extracted aspect $a \in \mathcal{A}^{\text{ext}}_q$ as $\mathbf{z}^{a}=g(a)$ and calculating $S_{a,d_{i,r}} = f(\mathbf{z}^{a},\mathbf{z}^{d_{i,r}})$. Then, aspect-item scores $S_{a,i} \in \mathbb{R}$ are obtained by aggregating the top $K_R$ aspect-review scores via Eq. \eqref{eqn:LF} with aspect-review scores instead of query-review scores. For each extracted aspect $a$, the top-$K_I$ scoring items are ordered into a list
\begin{align*}
L^a = (i_1,...,i_{K_I}) \quad  \mbox{s.t.} \; & i_1 \in \arg \max_i \{ S_{a,i}\} \\ & S_{a,i_k}, \geq S_{a,i_{k+1}}, \quad \forall i_{k} \in L^q.
\end{align*}

Figure \ref{fig:mono_vs_aspect}b) demonstrates aspect-item scoring and how it can alleviate the biases of standard LF. In this figure, the red and green points are embeddings of the reviews of item $i$ describing  aspect $a^{\text{item}}_{i,1}$ and $a^{\text{item}}_{i,2}$, respectively --- both these aspects are assumed to be relevant to the query. Though the former aspect is more frequent, an equal number ($K_R$) of reviews for each aspect will be used during score fusion --- as long as the aspect review embeddings are similar enough to the relevant query aspect embedding, and the total number of reviews for an aspect is at least $K_R$. In contrast, Figure \ref{fig:mono_vs_aspect}a) shows how standard (monolithic) LF will take a biased review sample of the first aspect since it is more frequently mentioned by reviews and $\mathbf{z}^q$ happens to be closer to those review embeddings. To differentiate between LF for RIR proposed by Abdollah Pour \textit{et al.}, and Aspect Fusion, we will refer to LF as \textit{Monolithic LF} since it uses the full query.

\subsubsection{Aspect-Item Score Fusion} After aspect-item scoring, we must aggregate the $A^e_q$ top $K_I$ item lists for each aspect $\{ L^a\}_{a_\in \mathcal{A}^{\text{ext}}_q}$ into a single ranked list of top-$K_I$ items for the query, $L^q$. We examine six aggregation strategies, which can be categorized as four score aggregation methods and two rank aggregation methods. The score-based variants convert the $A^e_q$ aspect-item scores into a query-item score $S_{q,i}$ using
\begin{enumerate}
\item \textbf{AMean}: Arithmatic mean
    \item \textbf{GMean}: Geometric mean
    \item \textbf{HMean}: Harmonic mean
    \item \textbf{Min}: Minimum
\end{enumerate}
to return the final ranked list $L^q$. The two rank-based list aggregation methods include:
\begin{enumerate}
    \item \textbf{Borda}: Borda count
    \item \textbf{R-R}: Round-robin (interleaved) merge.
\end{enumerate}

In Borda Count, the score for a given item $i$ is calculated as follows:
$\sum_{j=1}^{A^e_q} (K_I - \text{rank}^{L^{a_j}}_i + 1)$, where $\text{rank}^{L^{a_j}}_i$ is the rank of item $i$ in list $L^{a_j}$. In a round-robin merge of $A^e_q$ lists, elements from each list are merged in a cyclic order, and when a conflict arises with a particular item, that item is skipped and the merge continues from the same list. 

\subsection{LLM Reranking} 
In addition to Aspect Fusion, we also introduce an LLM reranking step for MA-RIR --- to the best of our knowledge LLM reranking has not been previously studied in a reviewed-item setting. Our goal is to understand whether LLMs in cross-encoder (CE) or ZS listwise \cite{ma2023zeroshot} settings can fuse reviews of multi-aspect items for effective reranking. 

After a list $L^q$ of top $K_I$ items is returned from the first stage, $K_R$ reviews for each item need to be given to the LLM for what we call \textit{fusion-during-reranking}. For Monolithic LF, these $K_R$ reviews are simply the $K_R$ reviews used for LF. For Aspect Fusion, since $K_R$ reviews were used for fusion \textit{with each aspect,} we propose to perform a round-robin merge of the top $K_R$ review lists for each aspect in order to preserve a balanced distribution of reviews across aspects. 

For a CE, reviews are simply concatenated and cross-encoded with the query. For listwise reranking, our prompt provides the LLM with the query, initial ranked list of item IDs, reviews for each item, and instructions to order the items based on relevance to the query --- the full listwise reranking prompt is in Figure \ref{fig:lw_rerank_prompt} in the Appendix.

\section{Experimental Method}

We perform simulations on generated review data to study the effect of aspect balance across reviews and test our hypothesis that Aspect Fusion is more robust to aspect imbalances than Monolithic LF. While using synthetic data exposes our results to biases from the data generation process, we are able to generate synthetic review distributions with far greater control that would have been possible several years ago before the advent of LLMs. We specifically design experiments to study the performance of Aspect Fusion vs Monolithic LF under the presence of aspect imbalance, both in the form of disjointedness of aspects across reviews and imbalanced aspect frequencies. 

In order to perform our experimentation, we need a dataset that has (a) multi-aspect queries and items, (b) GT aspect labels and (c) item reviews. To the best of our knowledge, there is no existing dataset with all of these properties. However, the recently-released Recipe-MPR dataset \cite{10.1145/3539618.3591880} includes properties (a) and (b). We leverage this dataset and generate item reviews using GPT-4. 

\begin{table}[h]
\caption{Distribution of ground truth (GT) and LLM-extracted aspects for Recipe-MPR queries and  items
\label{tab:aspect_dist}}
\footnotesize
\begin{tabular}{lrrrrrrrr}
\toprule
\# of Aspects & 1 & 2 & 3 & 4 & 5 & 6 & 7 & 8 \\
\midrule
Items (GT) & 76 & 282 & 72 & 29 & 10 & 1 & 2 & 1 \\
\midrule
Queries (GT) & 0 & 294 & 103 & 14 & 0 & 0 & 0 & 0 \\
\midrule
Queries (Extracted) & 0 & 2 & 342 & 55 & 12 & 0 & 0 & 0 \\
\bottomrule
\end{tabular}
\end{table}

\subsection{Data Generation}\label{sec:data_gen}
We create four datasets for our experiments based on the Recipe-MPR dataset and our new LLM-generated reviews.

Firstly, the \textit{fully overlapping} dataset includes 20 reviews per item, which each mention all of the aspects of the item. Secondly, the \textit{fully disjoint} dataset includes 10 reviews for each aspect of a given item. We also modify the fully disjoint dataset to create two datasets with imbalanced aspect frequencies. In the \textit{one rare aspect} dataset, we remove all but one of the reviews for a randomly-selected aspect of each item. In the \textit{one popular aspect} dataset, we keep all ten reviews for only one randomly-selected aspect of each item, and keep only one review for the other aspects.

In order to generate reviews, the GT aspects for each correct item in Recipe-MPR were used to prompt GPT-4. The total number of items for which there were GT aspects is 473. The distribution of the number of aspects per query and item is shown in Table \ref{tab:aspect_dist}. On average, each item has 2.2 aspects. The prompts we used to generate the reviews are included in the Appendix.

Recipe-MPR contains logical AND queries with ground truth (GT) labels for the query aspects. Refer to subsection \ref{ssec:multi_asp_queries} for an example of a query $q$ and its GT aspects, $\mathcal{A}^{\text{query}}_q$. Since the focus of this paper is on MA-RIR, we only included the 411 queries whose associated correct item had at least two aspects. For each of these queries, we used two-shot examples to have GPT-4 extract ``at least two non-overlapping spans'' representing the relevant aspects in the query.

\subsection{Experimental Details}
For our query and review embeddings, we used TAS-B \cite{hofstatter2021efficiently}. For the listwise reranking experiments, we used the gpt-3.5-turbo-16k model. For the CE reranking experiments, the model used was ms-marco-MiniLM-L-12-v2 \footnote{https://huggingface.co/cross-encoder/ms-marco-MiniLM-L-12-v2}. 

\section{Experimental Results}

\setlength{\tabcolsep}{3pt} 

\begin{figure}
    \centering
    \includegraphics[width=\linewidth]{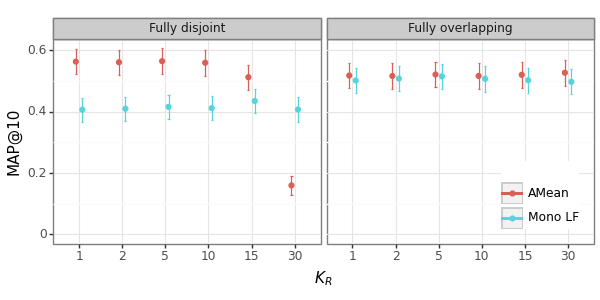}
    \caption{Monolithic LF versus Aspect Fusion with AMean aggregation. Both methods perform similarly on the fully overlapping dataset, but Aspect Fusion performs significantly better than Monolithic LF for the fully disjoint dataset and $K_R < 30$. For the fully disjoint dataset, Aspect Fusion drops in performance for $K_R > 10$ because when $K_R$ exceeds the number of reviews per aspect, scoring is based on reviews that are irrelevant to the given aspect. This decline in performance does not apply in the fully overlapping case.}
    \label{fig:rq1}
\end{figure}

\begin{table*}[h!]
\caption{LF versus Aspect Fusion with six various aggregation functions for both the Fully Disjoint and Fully Overlapping datasets with 95\% error margins in parentheses.}
\label{tab:rq1}
\centering{}
\footnotesize
\setlength\tabcolsep{1.5pt}
\renewcommand{\arraystretch}{1.2} 
\begin{tabular}{|l|l|cc|cc|cc|cc|cc|cc|}
\hline
 & $K_R$ & \multicolumn{2}{|c|}{1} & \multicolumn{2}{|c|}{2} & \multicolumn{2}{|c|}{5} & \multicolumn{2}{|c|}{10} & \multicolumn{2}{|c|}{15} & \multicolumn{2}{|c|}{30} \\
\hline
 Dataset &   & MAP@10 & Re@10 & MAP@10 & Re@10 & MAP@10 & Re@10 & MAP@10 & Re@10 & MAP@10 & Re@10 & MAP@10 & Re@10 \\
\hline
\multirow{7}{*}{\shortstack[l]{Fully\\disjoint}} & Mono LF & .41 (.04) & .67 (.05) & .41 (.04) & .69 (.04) & .42 (.04) & .69 (.04) & .41 (.04) & .70 (.04) & .43 (.04) & .70 (.04) & \textbf{.41 (.04)} & \textbf{.68 (.05)} \\ \cdashline{2-14} & AMean & .56 (.04) & \textbf{.77 (.04)} & .56 (.04) & \textbf{.77 (.04)} & .56 (.04) & \textbf{.78 (.04)} & \textbf{.56 (.04)} & .76 (.04) & .51 (.04) & \textbf{.75 (.04)} & .16 (.03) & .24 (.04) \\
 & Borda & .38 (.04) & .56 (.05) & .39 (.04) & .59 (.05) & .38 (.04) & .58 (.05) & .37 (.04) & .57 (.05) & .35 (.04) & .53 (.05) & .13 (.03) & .21 (.04) \\
 & GMean & \textbf{.57 (.04)} & \textbf{.77 (.04)} & .56 (.04) & \textbf{.77 (.04)} & .56 (.04) & \textbf{.78 (.04)} & \textbf{.56 (.04)} & \textbf{.77 (.04)} & .51 (.04) & \textbf{.75 (.04)} & .16 (.03) & .24 (.04) \\
 & HMean & \textbf{.57 (.04)} & \textbf{.77 (.04)} & \textbf{.57 (.04)} & \textbf{.77 (.04)} & \textbf{.57 (.04)} & \textbf{.78 (.04)} & \textbf{.56 (.04)} & \textbf{.77 (.04)} & \textbf{.52 (.04)} & \textbf{.75 (.04)} & .16 (.03) & .24 (.04) \\
 & Min & .43 (.04) & .62 (.05) & .01 (.01) & .03 (.02) & .01 (.01) & .03 (.02) & .01 (.01) & .03 (.02) & .01 (.01) & .03 (.02) & .01 (.01) & .03 (.02) \\
 & R-R & .21 (.03) & .66 (.05) & .21 (.03) & .66 (.05) & .21 (.03) & .66 (.05) & .21 (.03) & .66 (.05) & .19 (.02) & .66 (.05) & .06 (.02) & .22 (.04) \\
\cline{1-14}
\multirow{7}{*}{\shortstack[l]{Fully\\overlapping}} & Mono LF & .50 (.04) & .73 (.04) & .51 (.04) & .75 (.04) & .51 (.04) & .74 (.04) & .51 (.04) & .75 (.04) & .50 (.04) & .75 (.04) & .50 (.04) & .75 (.04) \\ \cdashline{2-14} & AMean & \textbf{.52 (.04)} & .74 (.04) & \textbf{.52 (.04)} & \textbf{.76 (.04)} & \textbf{.52 (.04)} & \textbf{.77 (.04)} & \textbf{.52 (.04)} & .75 (.04) & .52 (.04) & .75 (.04) & \textbf{.53 (.04)} & .74 (.04) \\
 & Borda & .33 (.04) & .51 (.05) & .33 (.04) & .52 (.05) & .34 (.04) & .52 (.05) & .34 (.04) & .53 (.05) & .33 (.04) & .52 (.05) & .33 (.04) & .51 (.05) \\
 & GMean & \textbf{.52 (.04)} & \textbf{.75 (.04)} & \textbf{.52 (.04)} & \textbf{.76 (.04)} & \textbf{.52 (.04)} & \textbf{.77 (.04)} & \textbf{.52 (.04)} & \textbf{.76 (.04)} & .52 (.04) & \textbf{.76 (.04)} & \textbf{.53 (.04)} & .75 (.04) \\
 & HMean & \textbf{.52 (.04)} & \textbf{.75 (.04)} & \textbf{.52 (.04)} & \textbf{.76 (.04)} & \textbf{.52 (.04)} & .76 (.04) & \textbf{.52 (.04)} & \textbf{.76 (.04)} & \textbf{.53 (.04)} & \textbf{.76 (.04)} & .52 (.04) & \textbf{.76 (.04)} \\
 & Min & .32 (.04) & .52 (.05) & .01 (.01) & .03 (.02) & .01 (.01) & .03 (.02) & .01 (.01) & .03 (.02) & .01 (.01) & .03 (.02) & .01 (.01) & .03 (.02) \\
 & R-R & .17 (.02) & .61 (.05) & .17 (.02) & .60 (.05) & .17 (.02) & .61 (.05) & .18 (.02) & .63 (.05) & .18 (.02) & .63 (.05) & .17 (.02) & .62 (.05) \\
\cline{1-14}
\end{tabular}
\end{table*}

\subsubsection*{\textbf{RQ1: Is Aspect Fusion helpful when item aspects are discussed disjointly across reviews?}}

\hfill

Table \ref{tab:rq1} lists the mean absolute precision at 10 (MAP@10) and recall@10 (Re@10) of the stage 1 dense retrieval for various settings of $K_R$. The table is broken up according to whether the disjoint or overlapping reviews are used. Throughout this paper, we show results for $K_I = 10$. In our experiments we noticed that varying $K_I$ led to minor changes in the results. For completeness, we report results for $K_I = 5$ in the Appendix.

We see that for the fully overlapping dataset, Aspect Fusion is approximately equivalent to the Monolithic LF approach, while for the fully disjoint dataset, Aspect Fusion score aggregation approaches (arithmetic mean, harmonic mean, and geometric mean) offer a significant improvement in performance compared to the Monolithic LF approach. This pattern offers empirical evidence that Aspect Fusion is better suited to disjoint aspect distributions than Monolithic LF. More specifically, this suggests that Monolithic LF is not symmetrical across aspects, and fails to consider information from each of the aspects in a balanced way.

Additionally, for the fully disjoint dataset, the performance of the aspect-based approach suffers for $K_R > 10$. This can be explained by the fact that when $K_R$ exceeds the number of disjoint reviews available for a given aspect (10 in this data), the aspect-based methods will score items based on reviews that are irrelevant to a given aspect. This could result in correct items receiving low scores for some aspects. We conclude that Aspect Fusion should use $K_R \leq R_i^{a,\text{min}}$, where $R_i^{a,\text{min}}$ is the smallest number of reviews for an item $i$ for an aspect, in order to avoid this performance drop. 

Furthermore, the fact that the score aggregation methods outperform the rank-based aggregation methods (R-R and Borda) offers evidence that the embedding similarity scores contain significant information about how well an item's reviews align with a given query aspect, above and beyond that item's rank relative to the other candidate items. Considering the simplicity and strong performance of AMean score aggregation, we focus on this Aspect Fusion method in the remaining results below.

\begin{table*}
\caption{Effect of aspect frequency imbalance on Monolithic LF and Aspect Fusion. ``Balanced frequency'' refers to the fully disjoint dataset where all item aspects have the same number of reviews. The values in parentheses indicate the 95\% error margin.}
\label{tab:rq3}
\footnotesize
\centering{}
\setlength\tabcolsep{1.5pt}
\renewcommand{\arraystretch}{1.2} 
\begin{tabular}{|l|l|cc|cc|cc|cc|cc|cc|}
\hline
 \multirow{2}{*}{Dataset} & $K_R$ & \multicolumn{2}{|c|}{1} & \multicolumn{2}{|c|}{2} & \multicolumn{2}{|c|}{5} & \multicolumn{2}{|c|}{10} & \multicolumn{2}{|c|}{15} & \multicolumn{2}{|c|}{30} \\
 &   & MAP@10 & Re@10 & MAP@10 & Re@10 & MAP@10 & Re@10 & MAP@10 & Re@10 & MAP@10 & Re@10 & MAP@10 & Re@10 \\
\hline
\multirow{2}{*}{\shortstack[l]{Balanced\\Frequency}} & AMean & \textbf{.56 (.04)} & \textbf{.77 (.04)} & \textbf{.56 (.04)} & \textbf{.77 (.04)} & \textbf{.56 (.04)} & \textbf{.78 (.04)} & \textbf{.56 (.04)} & \textbf{.76 (.04)} & \textbf{.51 (.04)} & \textbf{.75 (.04)} & .16 (.03) & .24 (.04) \\
 & Mono LF & .41 (.04) & .67 (.05) & .41 (.04) & .69 (.04) & .42 (.04) & .69 (.04) & .41 (.04) & .70 (.04) & .43 (.04) & .70 (.04) & \textbf{.41 (.04)} & \textbf{.68 (.05)} \\
\cline{1-14}
\multirow{2}{*}{\shortstack[l]{One Popular\\Aspect}} & AMean & \textbf{.52 (.04)} & \textbf{.73 (.04)} & \textbf{.43 (.04)} & \textbf{.68 (.05)} & \textbf{.33 (.04)} & \textbf{.58 (.05)} & \textbf{.27 (.04)} & \textbf{.52 (.05)} & .02 (.01) & .03 (.02) & .01 (.01) & .03 (.02) \\
 & Mono LF & .36 (.04) & .62 (.05) & .32 (.04) & .60 (.05) & .28 (.04) & .53 (.05) & .26 (.04) & .49 (.05) & \textbf{.27 (.04)} & \textbf{.51 (.05)} & \textbf{.27 (.04)} & \textbf{.52 (.05)} \\
\cline{1-14}
\multirow{2}{*}{\shortstack[l]{One Rare\\Aspect}} & AMean & \textbf{.52 (.04)} & \textbf{.74 (.04)} & \textbf{.45 (.04)} & \textbf{.67 (.05)} & \textbf{.36 (.04)} & \textbf{.58 (.05)} & \textbf{.34 (.04)} & \textbf{.54 (.05)} & .15 (.03) & .23 (.04) & .06 (.02) & .09 (.03) \\
 & Mono LF & .39 (.04) & .65 (.05) & .36 (.04) & .64 (.05) & .32 (.04) & .57 (.05) & .30 (.04) & .53 (.05) & \textbf{.32 (.04)} & \textbf{.55 (.05)} & \textbf{.29 (.04)} & \textbf{.51 (.05)} \\
\cline{1-14}
\end{tabular}
\end{table*}

\begin{figure}
    \centering
    \includegraphics[width=\linewidth]{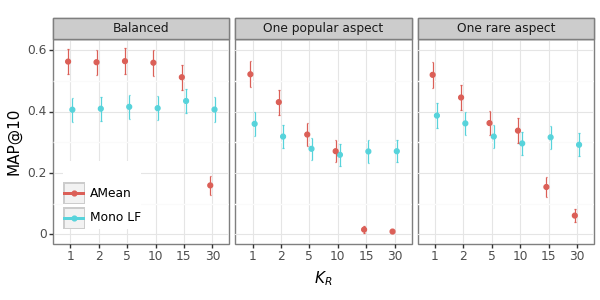}
    \caption{Effect of Aspect Frequency. Aspect Fusion performs better than Monolithic LF for low values of $K_R$, but suffers for higher values of $K_R$. This pattern is explained in the discussion of RQ1.}
    \label{fig:rq3}
\end{figure}

\subsubsection*{\textbf{RQ2: How does review aspect frequency imbalance affect Monlithic LF and Aspect Fusion?}}

\hfill

Table \ref{tab:rq3} shows the performance of the stage 1 dense retrieval for the balanced frequency (fully disjoint) dataset and the two datasets with imbalance in the review aspect frequency. These results are also presented visually in Figure \ref{fig:rq3}. Note that this imbalance can only be analyzed for the case where the reviews cover disjoint, rather than overlapping, aspects.

Based on our conclusion above, we focus on the results for $K_R = 1$ in this section, since for the datasets with imbalanced review aspect frequency, $R_i^{a,\text{min}} = 1$. We see that there is a significant decrease in performance for all methods when aspect frequency imbalance is introduced. This result suggests that balance in reviews across aspects is helpful for both Monolithic LF and Aspect Fusion. 

Furthermore, for $K_R = 1$, the performance of Monolithic LF decreases more when aspect frequency imbalance is introduced, compared to for Aspect Fusion methods. For example, the MAP@10 of Monolithic LF decreased from 0.41 to 0.36 on the \textit{one popular aspect} dataset, representing a 12\% drop, compared to a 7\% drop for the Aspect Fusion approach. This suggests Aspect Fusion methods may be more robust to aspect frequency imbalance.

Lastly, we note that the performance of Monolithic LF decreases as $K_R$ grows large, which occurs because any relevant item aspects that are infrequently reviewed (there is only 1 review for rare aspects in these datasets) will contribute less and less to the query-item score with an increase in $K_R$.  

\begin{table*}
\caption{Effect of using LLM extracted query aspects vs. GT query aspects on Monolithic LF and Aspect Fusion. The values in parentheses indicate the 95\% error margin.}
\label{tab:rq4}
\footnotesize
\centering{}
\setlength\tabcolsep{1.5pt}
\renewcommand{\arraystretch}{1.2} 
\begin{tabular}{|l|l|cc|cc|cc|cc|cc|cc|}
\hline
 & $K_R$ & \multicolumn{2}{|c|}{1} & \multicolumn{2}{|c|}{2} & \multicolumn{2}{|c|}{5} & \multicolumn{2}{|c|}{10} & \multicolumn{2}{|c|}{15} & \multicolumn{2}{|c|}{30} \\
 &   & MAP@10 & Re@10 & MAP@10 & Re@10 & MAP@10 & Re@10 & MAP@10 & Re@10 & MAP@10 & Re@10 & MAP@10 & Re@10 \\
\hline
\multirow{2}{*}{\shortstack[l]{Extracted\\query aspects}} & AMean & \textbf{.46 (.04)} & \textbf{.70 (.04)} & \textbf{.47 (.04)} & \textbf{.72 (.04)} & \textbf{.47 (.04)} & \textbf{.72 (.04)} & \textbf{.46 (.04)} & \textbf{.71 (.04)} & \textbf{.46 (.04)} & \textbf{.71 (.04)} & .15 (.03) & .23 (.04) \\
 & Mono LF & .41 (.04) & .67 (.05) & .41 (.04) & .69 (.04) & .42 (.04) & .69 (.04) & .41 (.04) & .70 (.04) & .43 (.04) & .70 (.04) & \textbf{.41 (.04)} & \textbf{.68 (.05)} \\
\cline{1-14}
\multirow{2}{*}{\shortstack[l]{GT query\\aspects}} & AMean & \textbf{.56 (.04)} & \textbf{.77 (.04)} & \textbf{.56 (.04)} & \textbf{.77 (.04)} & \textbf{.56 (.04)} & \textbf{.78 (.04)} & \textbf{.56 (.04)} & \textbf{.76 (.04)} & \textbf{.51 (.04)} & \textbf{.75 (.04)} & .16 (.03) & .24 (.04) \\
 & Mono LF & .41 (.04) & .67 (.05) & .41 (.04) & .69 (.04) & .42 (.04) & .69 (.04) & .41 (.04) & .70 (.04) & .43 (.04) & .70 (.04) & \textbf{.41 (.04)} & \textbf{.68 (.05)} \\
\cline{1-14}
\end{tabular}
\end{table*}

\begin{figure}
    \centering
    \includegraphics[width=\linewidth]{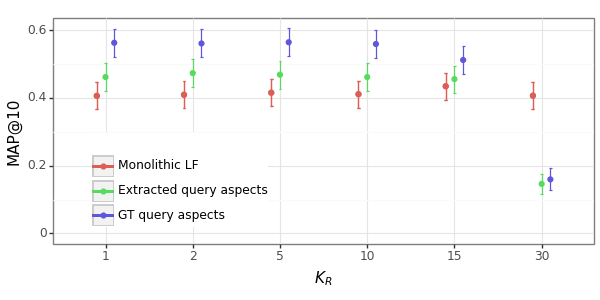}
    \caption{Aspect Fusion with GT vs extracted query aspects with fully disjoint reviews. Although GT query aspects perform better, Aspect Fusion still offers an improvement over Monolithic LF with extracted query aspects.}
    \label{fig:rq4}
\end{figure}

\subsubsection*{\textbf{RQ3: How does the use of extracted query aspects instead of GT query aspects affect Aspect Fusion?}}

\hfill

Table \ref{tab:rq4} shows the same results as Table \ref{tab:rq1} except for the case of the extracted query aspects. These results are also presented visually in Figure \ref{fig:rq4}. At $K_R = 1$, while the MAP@10 of Aspect Fusion drops from 0.56 with GT aspects to 0.46 with extracted aspects, it remains higher than the 0.41 MAP@10 of Monolithic LF. This result implies that Aspect Fusion is useful even when GT query aspects are unknown.


\begin{table*}
\caption{Reranker performance of CE and LW LLMs for various $K_R$ values. ``No'' refers to the case where no reranking is applied, and is equivalent to the stage 1 results. The values in parentheses indicate the 95\% error margin.}
\label{tab:rq2}
\footnotesize
\centering{}
\setlength\tabcolsep{1.5pt}
\renewcommand{\arraystretch}{1.2} 
\begin{tabular}{|l|l|l|cc|cc|cc|cc|cc|cc|}
\hline
 &  & $K_R$ & \multicolumn{2}{|c|}{1} & \multicolumn{2}{|c|}{2} & \multicolumn{2}{|c|}{5} & \multicolumn{2}{|c|}{10} & \multicolumn{2}{|c|}{15} & \multicolumn{2}{|c|}{30} \\
 &  &   & MAP@10 & Re@10 & MAP@10 & Re@10 & MAP@10 & Re@10 & MAP@10 & Re@10 & MAP@10 & Re@10 & MAP@10 & Re@10 \\
\hline
\multirow{6}{*}{\shortstack[l]{Fully\\disjoint}} & \multirow{3}{*}{AMean} & CE & .36 (.04) & .77 (.04) & .51 (.04) & .77 (.04) & .53 (.04) & .78 (.04) & .53 (.04) & .76 (.04) & \textbf{.53 (.04)} & .75 (.04) & \textbf{.16 (.03)} & .24 (.04) \\
 &  & LW & .40 (.04) & .77 (.04) & .45 (.04) & .77 (.04) & .53 (.04) & .78 (.04) & .55 (.04) & .76 (.04) & .52 (.04) & .75 (.04) & \textbf{.16 (.03)} & .24 (.04) \\
 &  & No & \textbf{.56 (.04)} & .77 (.04) & \textbf{.56 (.04)} & .77 (.04) & \textbf{.56 (.04)} & .78 (.04) & \textbf{.56 (.04)} & .76 (.04) & .51 (.04) & .75 (.04) & \textbf{.16 (.03)} & .24 (.04) \\
\cline{2-15}
 & \multirow{3}{*}{Mono LF} & CE & .35 (.04) & .67 (.05) & .38 (.04) & .69 (.04) & .40 (.04) & .69 (.04) & \textbf{.44 (.04)} & .70 (.04) & \textbf{.47 (.04)} & .70 (.04) & \textbf{.47 (.04)} & .68 (.05) \\
 &  & LW & .33 (.04) & .67 (.05) & .34 (.04) & .69 (.04) & .38 (.04) & .69 (.04) & .40 (.04) & .70 (.04) & .42 (.04) & .70 (.04) & .46 (.04) & .68 (.05) \\
 &  & No & \textbf{.41 (.04)} & .67 (.05) & \textbf{.41 (.04)} & .69 (.04) & \textbf{.42 (.04)} & .69 (.04) & .41 (.04) & .70 (.04) & .43 (.04) & .70 (.04) & .41 (.04) & .68 (.05) \\
\cline{1-15} \cline{2-15}
\multirow{6}{*}{\shortstack[l]{Fully\\overlapping}} & \multirow{3}{*}{AMean} & CE & .51 (.04) & .74 (.04) & \textbf{.52 (.04)} & .76 (.04) & .52 (.04) & .77 (.04) & .51 (.04) & .75 (.04) & .48 (.04) & .75 (.04) & .50 (.04) & .74 (.04) \\
 &  & LW & .43 (.04) & .74 (.04) & .48 (.04) & .76 (.04) & \textbf{.55 (.04)} & .77 (.04) & \textbf{.53 (.04)} & .75 (.04) & \textbf{.53 (.04)} & .75 (.04) & .52 (.04) & .74 (.04) \\
 &  & No & \textbf{.52 (.04)} & .74 (.04) & \textbf{.52 (.04)} & .76 (.04) & .52 (.04) & .77 (.04) & .52 (.04) & .75 (.04) & .52 (.04) & .75 (.04) & \textbf{.53 (.04)} & .74 (.04) \\
\cline{2-15}
 & \multirow{3}{*}{Mono LF} & CE & \textbf{.50 (.04)} & .73 (.04) & .50 (.04) & .75 (.04) & .50 (.04) & .74 (.04) & .50 (.04) & .75 (.04) & .48 (.04) & .75 (.04) & .50 (.04) & .75 (.04) \\
 &  & LW & .45 (.04) & .73 (.04) & .47 (.04) & .75 (.04) & \textbf{.52 (.04)} & .74 (.04) & \textbf{.53 (.04)} & .75 (.04) & \textbf{.53 (.04)} & .75 (.04) & \textbf{.52 (.04)} & .75 (.04) \\
 &  & No & \textbf{.50 (.04)} & .73 (.04) & \textbf{.51 (.04)} & .75 (.04) & .51 (.04) & .74 (.04) & .51 (.04) & .75 (.04) & .50 (.04) & .75 (.04) & .50 (.04) & .75 (.04) \\
\cline{1-15} \cline{2-15}
\end{tabular}
\end{table*}

\begin{figure}
    \centering
    \includegraphics[width=\linewidth]{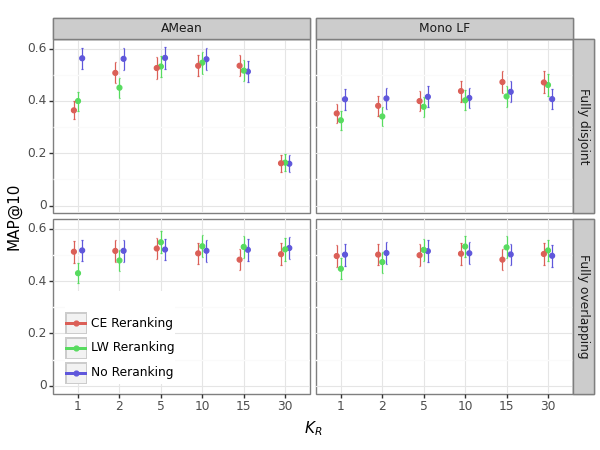}
    \caption{Comparison of reranking methods. Performance generally increases as more reviews are included in the LLM --- using too few reviews can hurt performance.}
    \label{fig:rq2}
\end{figure}

\begin{figure}
    \centering
    \includegraphics[width=\linewidth]{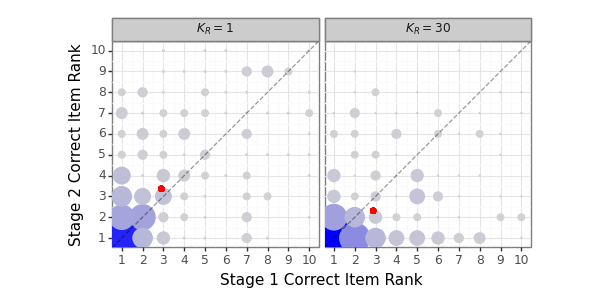}
    \caption{Ranks of correct items after Stage 1 Monolithic LF (x axis) and Stage 2 Cross-Encoder reranking (y  axis) on the Fully Disjoint dataset. Circle size is proportional to position frequency, and the center of mass is shown in red. For $K_R = 1$, most of the mass lies above the diagonal line, meaning the reranker has worsened performance. On the other hand, for $K_R = 30$, most of the mass lies below the diagonal line, meaning that the reranker has improved the performance.}
    \label{fig:rq2_bubble}
\end{figure}
\subsubsection*{\textbf{RQ4: Are LLMs effective MA-RIR rerankers?}}

\hfill

Table \ref{tab:rq2} summarizes the performance of the listwise\footnote{Approximately 1\% of queries had only 9 items returned by the listwise reranker instead of 10 --- this was an error in generative retrieval.}
and cross-encoder rerankers. We see there is a beneficial effect to increasing the number of reviews $K_R$ given to the language model for both CE and listwise reranking. Specifically, for reranking Monolithic LF on the fully disjoint dataset, listwise MAP@10 improves from 0.33 to 0.46, for $K_R=1$ and $K_R=30$, respectively. Similarly, CE MAP@10 improves from 0.35 MAP@10 at $K_R = 1$ to 0.47 at $K_R = 30$. We conjecture this large increase in MAP@10 with $K_R$ is due to the quadratic nature of cross-attention across input text. 

Since Aspect Fusion did best with low $K_R$ values, a possible reason that we did not observe any benefits of LLM reranking for Aspect Fusion is because $K_R$ was not high enough. Also, while some reranking settings showed 2nd stage MAP@10 increases over 1st stage values (such as at $K_R = 30$ reranking of Monolithic LF for fully disjoint data), when too few reviews were given to the reranker, the second stage sometimes made performance worse, such as at $K_R =1$. 

Figure \ref{fig:rq2_bubble} shows a heatmap of the ranks assigned to the correct items by the stage 1 retriever and stage 2 reranker. An effective reranker would consistently improve the ranks for the correct item, and this would result in the center of mass lying below the anti-diagonal. We see that this is indeed the case for a high value of $K_R$, but is not the case for a low value of $K_R$. The raw values underlying this figure are provided in the Appendix.


\section{Related Work}\label{sec:rel_work}

\subsection{Multi-level Retrieval}\label{sec:multi-level retrieval}
The most relevant work to ours is that on RIR by Abdollah Pour \textit{et al.} \cite{abdollah2023self}, which formulates the RIR problem and studies EF and LF approaches. In addition to LF with an off-the-shelf bi-encoder such as TAS-B, the authors also contrastively fine tune an encoder for LF and show performance improvements over off-the-shelf LF. Extending their contrastive learning approach to MA-RIR Aspect Fusion is a natural direction for future work. As mentioned in Section \ref{sec:background}, Zhang and Balog \cite{zhang2017design} have previously studied the Object Fusion problem, which allows for more general two-level structures than RIR (in which a low-level document cannot describe more than one high-level object). However, they did not study neural techniques or multi-aspect retrieval, which are key to our work.

\subsection{Multi-aspect Retrieval}
In addition to releasing Recipe-MPR, which was used to generate review distributions in this work, Zhang \textit{et al} \cite{10.1145/3539618.3591880} use the queries and items in Recipe-MPR in a multi-aspect question-answering setting, and find that FS GPT-3 listwise prompting achieves far superior accuracy to all other methods. However, it is computationally infeasible to use such listwise prompting methods for first stage retrieval. Kong \textit{et al.} \cite{kong2022multi} consider multiple aspects when calculating relevance scores in dense retrieval, but assume documents and queries contain a fixed number of aspects from known categories. Similarly, the label aggregation method of Kang \textit{et al.} \cite{kang2012learning} explicitly deals with multiple query aspects, but has fixed number of known categories.

Another methods called Multi-Aspect Dense Retrieval (MADRM) \cite{kong2022multi} learns early fusion embeddings of documents and queries by extracting and then aggregating their aspects, and report improvements over Monolithic LF baselines. DORIS-MAE \cite{wang2023dorismae} presents a dataset that deconstructs complex queries into hierarchies of aspects and sub-aspects. Unlike our aspect extraction approach, which extracts aspects from queries using few-shot prompting with an LLM, DORIS-MAE predefines these aspects and their corresponding topic hierarchy for both queries and document corpora.

Finally, some recent works study multi-aspect LLM-driven conversational recommendation \cite{deldjoo2024review}, including work on preference elicitation over multiple aspects \cite{austin2024bayesian} and knowledge graph based topic-guided chatbots \cite{zhou2020towards}.


\section{Conclusions}
By extending  reviewed-item-retrieval (RIR) to a setting with multi-aspect queries and items, we were able to both theoretically and empirically demonstrate the failure modes of Monolithic Late Fusion (LF) when there is an imbalance in how aspects are distributed across reviews. Specifically, since Monolithic LF is aspect-agnostic, it is subject to a frequency bias in its review selection towards more popular aspects. Furthermore, the disjointedness of aspects across reviews can induce a selection bias towards certain aspects if monolithic multi-aspect query embeddings are closer to review embeddings for those aspects.

To address these failure modes, we propose Aspect Fusion as a robust MA-RIR method for imbalanced review distributions. Using the recently released Recipe-MPR dataset, specifically designed to study multi-aspect retrieval, we design four generated datasets that allow us to empircally test the effects of review imbalances from aspect frequency and disjointess. Our experiments show that Aspect Fusion is much more robust to non-uniform review variations than Monolithic LF, outperforming the later with a 44\% MAP@10 increase on some distributions.  

\bibliography{refs}

\begin{thebibliography}{15}
\expandafter\ifx\csname natexlab\endcsname\relax\def\natexlab#1{#1}\fi
\providecommand{\url}[1]{\texttt{#1}}
\providecommand{\href}[2]{#2}
\providecommand{\path}[1]{#1}
\providecommand{\DOIprefix}{doi:}
\providecommand{\ArXivprefix}{arXiv:}
\providecommand{\URLprefix}{URL: }
\providecommand{\Pubmedprefix}{pmid:}
\providecommand{\doi}[1]{\href{http://dx.doi.org/#1}{\path{#1}}}
\providecommand{\Pubmed}[1]{\href{pmid:#1}{\path{#1}}}
\providecommand{\bibinfo}[2]{#2}
\ifx\xfnm\relax \def\xfnm[#1]{\unskip,\space#1}\fi
\bibitem[{Abdollah~Pour et~al.(2023)Abdollah~Pour, Farinneya, Toroghi, Korikov,
  Pesaranghader, Sajed, Bharadwaj, Mavrin, and Sanner}]{abdollah2023self}
\bibinfo{author}{M.~M. Abdollah~Pour}, \bibinfo{author}{P.~Farinneya},
  \bibinfo{author}{A.~Toroghi}, \bibinfo{author}{A.~Korikov},
  \bibinfo{author}{A.~Pesaranghader}, \bibinfo{author}{T.~Sajed},
  \bibinfo{author}{M.~Bharadwaj}, \bibinfo{author}{B.~Mavrin},
  \bibinfo{author}{S.~Sanner},
\newblock \bibinfo{title}{Self-supervised contrastive {BERT} fine-tuning for
  fusion-based reviewed-item retrieval},
\newblock in: \bibinfo{booktitle}{European Conference on Information
  Retrieval}, \bibinfo{organization}{Springer}, \bibinfo{year}{2023}, pp.
  \bibinfo{pages}{3--17}. \DOIprefix\doi{10.1007/978-3-031-28244-7_1}.
\bibitem[{Kemper et~al.(2024)Kemper, Cui, Dicarlantonio, Lin, Tang, Korikov,
  and Sanner}]{Kemper2024}
\bibinfo{author}{S.~Kemper}, \bibinfo{author}{J.~Cui},
  \bibinfo{author}{K.~Dicarlantonio}, \bibinfo{author}{K.~Lin},
  \bibinfo{author}{D.~Tang}, \bibinfo{author}{A.~Korikov},
  \bibinfo{author}{S.~Sanner},
\newblock \bibinfo{title}{Retrieval-augmented conversational recommendation
  with prompt-based semi-structured natural language state tracking},
\newblock in: \bibinfo{booktitle}{Proceedings of the 47th International ACM
  SIGIR Conference on Research and Development in Information Retrieval}, SIGIR
  '24, \bibinfo{publisher}{Association for Computing Machinery},
  \bibinfo{address}{New York, NY, USA}, \bibinfo{year}{2024}.
  \DOIprefix\doi{10.1145/3626772.3657670}.
\bibitem[{Zhang et~al.(2023)Zhang, Korikov, Farinneya, Abdollah~Pour,
  Bharadwaj, Pesaranghader, Huang, Lok, Wang, Jones, and
  Sanner}]{10.1145/3539618.3591880}
\bibinfo{author}{H.~Zhang}, \bibinfo{author}{A.~Korikov},
  \bibinfo{author}{P.~Farinneya}, \bibinfo{author}{M.~M. Abdollah~Pour},
  \bibinfo{author}{M.~Bharadwaj}, \bibinfo{author}{A.~Pesaranghader},
  \bibinfo{author}{X.~Y. Huang}, \bibinfo{author}{Y.~X. Lok},
  \bibinfo{author}{Z.~Wang}, \bibinfo{author}{N.~Jones},
  \bibinfo{author}{S.~Sanner},
\newblock \bibinfo{title}{Recipe-{MPR}: A test collection for evaluating
  multi-aspect preference-based natural language retrieval},
\newblock in: \bibinfo{booktitle}{Proceedings of the 46th International ACM
  SIGIR Conference on Research and Development in Information Retrieval}, SIGIR
  '23, \bibinfo{publisher}{Association for Computing Machinery},
  \bibinfo{address}{New York, NY, USA}, \bibinfo{year}{2023}, p.
  \bibinfo{pages}{2744–2753}. \DOIprefix\doi{10.1145/3539618.3591880}.
\bibitem[{Reimers and Gurevych(2019)}]{reimers2019sentence}
\bibinfo{author}{N.~Reimers}, \bibinfo{author}{I.~Gurevych},
\newblock \bibinfo{title}{Sentence-{BERT}: Sentence embeddings using {S}iamese
  {BERT}-networks},
\newblock in: \bibinfo{editor}{K.~Inui}, \bibinfo{editor}{J.~Jiang},
  \bibinfo{editor}{V.~Ng}, \bibinfo{editor}{X.~Wan} (Eds.),
  \bibinfo{booktitle}{Proceedings of the 2019 Conference on Empirical Methods
  in Natural Language Processing and the 9th International Joint Conference on
  Natural Language Processing (EMNLP-IJCNLP)}, \bibinfo{publisher}{Association
  for Computational Linguistics}, \bibinfo{address}{Hong Kong, China},
  \bibinfo{year}{2019}, pp. \bibinfo{pages}{3982--3992}.
  \DOIprefix\doi{10.18653/v1/D19-1410}.
\bibitem[{Johnson et~al.(2021)Johnson, Douze, and Jégou}]{johnson2019billion}
\bibinfo{author}{J.~Johnson}, \bibinfo{author}{M.~Douze},
  \bibinfo{author}{H.~Jégou},
\newblock \bibinfo{title}{Billion-scale similarity search with {GPU}s},
\newblock \bibinfo{journal}{IEEE Transactions on Big Data} \bibinfo{volume}{7}
  (\bibinfo{year}{2021}) \bibinfo{pages}{535--547}.
  \DOIprefix\doi{10.1109/TBDATA.2019.2921572}.
\bibitem[{Zhang and Balog(2017)}]{zhang2017design}
\bibinfo{author}{S.~Zhang}, \bibinfo{author}{K.~Balog},
\newblock \bibinfo{title}{Design patterns for fusion-based object retrieval},
\newblock in: \bibinfo{booktitle}{European Conference on Information
  Retrieval}, \bibinfo{organization}{Springer}, \bibinfo{year}{2017}, pp.
  \bibinfo{pages}{684--690}. \DOIprefix\doi{10.1007/978-3-319-56608-5_66}.
\bibitem[{Ethayarajh(2019)}]{ethayarajh-2019-contextual}
\bibinfo{author}{K.~Ethayarajh},
\newblock \bibinfo{title}{How contextual are contextualized word
  representations? {C}omparing the geometry of {BERT}, {ELM}o, and {GPT}-2
  embeddings},
\newblock in: \bibinfo{editor}{K.~Inui}, \bibinfo{editor}{J.~Jiang},
  \bibinfo{editor}{V.~Ng}, \bibinfo{editor}{X.~Wan} (Eds.),
  \bibinfo{booktitle}{Proceedings of the 2019 Conference on Empirical Methods
  in Natural Language Processing and the 9th International Joint Conference on
  Natural Language Processing (EMNLP-IJCNLP)}, \bibinfo{publisher}{Association
  for Computational Linguistics}, \bibinfo{address}{Hong Kong, China},
  \bibinfo{year}{2019}, pp. \bibinfo{pages}{55--65}. \URLprefix
  \url{https://aclanthology.org/D19-1006}.
  \DOIprefix\doi{10.18653/v1/D19-1006}.
\bibitem[{Ma et~al.(2023)Ma, Zhang, Pradeep, and Lin}]{ma2023zeroshot}
\bibinfo{author}{X.~Ma}, \bibinfo{author}{X.~Zhang},
  \bibinfo{author}{R.~Pradeep}, \bibinfo{author}{J.~Lin},
  \bibinfo{title}{Zero-shot listwise document reranking with a large language
  model}, \bibinfo{year}{2023}. \href{http://arxiv.org/abs/2305.02156}{{\tt
  arXiv:2305.02156}}.
\bibitem[{Hofst{\"a}tter et~al.(2021)Hofst{\"a}tter, Lin, Yang, Lin, and
  Hanbury}]{hofstatter2021efficiently}
\bibinfo{author}{S.~Hofst{\"a}tter}, \bibinfo{author}{S.-C. Lin},
  \bibinfo{author}{J.-H. Yang}, \bibinfo{author}{J.~Lin},
  \bibinfo{author}{A.~Hanbury},
\newblock \bibinfo{title}{Efficiently teaching an effective dense retriever
  with balanced topic aware sampling},
\newblock in: \bibinfo{booktitle}{Proceedings of the 44th International ACM
  SIGIR Conference on Research and Development in Information Retrieval},
  \bibinfo{year}{2021}, pp. \bibinfo{pages}{113--122}.
\bibitem[{Kong et~al.(2022)Kong, Khadanga, Li, Gupta, Zhang, Xu, and
  Bendersky}]{kong2022multi}
\bibinfo{author}{W.~Kong}, \bibinfo{author}{S.~Khadanga},
  \bibinfo{author}{C.~Li}, \bibinfo{author}{S.~K. Gupta},
  \bibinfo{author}{M.~Zhang}, \bibinfo{author}{W.~Xu},
  \bibinfo{author}{M.~Bendersky},
\newblock \bibinfo{title}{Multi-aspect dense retrieval},
\newblock in: \bibinfo{booktitle}{Proceedings of the 28th ACM SIGKDD Conference
  on Knowledge Discovery and Data Mining}, KDD '22,
  \bibinfo{publisher}{Association for Computing Machinery},
  \bibinfo{address}{New York, NY, USA}, \bibinfo{year}{2022}, p.
  \bibinfo{pages}{3178–3186}. \DOIprefix\doi{10.1145/3534678.3539137}.
\bibitem[{Kang et~al.(2012)Kang, Wang, Chang, and Tseng}]{kang2012learning}
\bibinfo{author}{C.~Kang}, \bibinfo{author}{X.~Wang},
  \bibinfo{author}{Y.~Chang}, \bibinfo{author}{B.~Tseng},
\newblock \bibinfo{title}{Learning to rank with multi-aspect relevance for
  vertical search},
\newblock in: \bibinfo{booktitle}{Proceedings of the Fifth ACM International
  Conference on Web Search and Data Mining}, WSDM '12,
  \bibinfo{publisher}{Association for Computing Machinery},
  \bibinfo{address}{New York, NY, USA}, \bibinfo{year}{2012}, p.
  \bibinfo{pages}{453–462}. \DOIprefix\doi{10.1145/2124295.2124350}.
\bibitem[{Wang et~al.(2024)Wang, Wang, Wang, Naidu, Bergen, and
  Paturi}]{wang2023dorismae}
\bibinfo{author}{J.~Wang}, \bibinfo{author}{K.~Wang},
  \bibinfo{author}{X.~Wang}, \bibinfo{author}{P.~Naidu},
  \bibinfo{author}{L.~Bergen}, \bibinfo{author}{R.~Paturi},
\newblock \bibinfo{title}{{DORIS-MAE}: Scientific document retrieval using
  multi-level aspect-based queries},
\newblock in: \bibinfo{booktitle}{Proceedings of the 37th International
  Conference on Neural Information Processing Systems}, NIPS '23,
  \bibinfo{publisher}{Curran Associates Inc.}, \bibinfo{address}{Red Hook, NY,
  USA}, \bibinfo{year}{2024}. \DOIprefix\doi{10.5555/3666122.3667790}.
\bibitem[{Deldjoo et~al.(2024)Deldjoo, He, McAuley, Korikov, Sanner, Ramisa,
  Vidal, Sathiamoorthy, Kasirzadeh, and Milano}]{deldjoo2024review}
\bibinfo{author}{Y.~Deldjoo}, \bibinfo{author}{Z.~He},
  \bibinfo{author}{J.~McAuley}, \bibinfo{author}{A.~Korikov},
  \bibinfo{author}{S.~Sanner}, \bibinfo{author}{A.~Ramisa},
  \bibinfo{author}{R.~Vidal}, \bibinfo{author}{M.~Sathiamoorthy},
  \bibinfo{author}{A.~Kasirzadeh}, \bibinfo{author}{S.~Milano},
\newblock \bibinfo{title}{A review of modern recommender systems using
  generative models (gen-recsys)},
\newblock in: \bibinfo{booktitle}{Proceedings of the 30th {ACM SIGKDD}
  Conference on Knowledge Discovery and Data Mining ({KDD} ’24), August
  25–29, 2024, Barcelona, Spain}, \bibinfo{year}{2024}.
\bibitem[{Austin et~al.(2024)Austin, Korikov, Toroghi, and
  Sanner}]{austin2024bayesian}
\bibinfo{author}{D.~E. Austin}, \bibinfo{author}{A.~Korikov},
  \bibinfo{author}{A.~Toroghi}, \bibinfo{author}{S.~Sanner},
\newblock \bibinfo{title}{Bayesian optimization with {LLM}-based acquisition
  functions for natural language preference elicitation},
\newblock in: \bibinfo{booktitle}{Proceedings of the 18th {ACM} Conference on
  Recommender Systems (Rec{S}ys'24)}, \bibinfo{year}{2024}.
\bibitem[{Zhou et~al.(2020)Zhou, Zhou, Zhao, Wang, and Wen}]{zhou2020towards}
\bibinfo{author}{K.~Zhou}, \bibinfo{author}{Y.~Zhou}, \bibinfo{author}{W.~X.
  Zhao}, \bibinfo{author}{X.~Wang}, \bibinfo{author}{J.-R. Wen},
\newblock \bibinfo{title}{Towards topic-guided conversational recommender
  system},
\newblock \bibinfo{journal}{arXiv preprint arXiv:2010.04125}
  (\bibinfo{year}{2020}).

\end{thebibliography}



\appendix

\section{Appendix A}

\subsection{LLM Prompts}

\begin{figure}[h]
    \centering
    \includegraphics[width=0.8\linewidth]{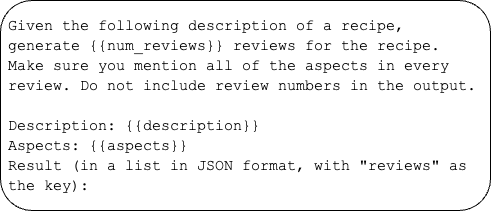}
    \caption{Overlapping Review Generation Prompt Used with GPT-4}
    \label{fig:overlapping_review_prompt}
\end{figure}

\begin{figure}[h]
    \centering
    \includegraphics[width=0.8\linewidth]{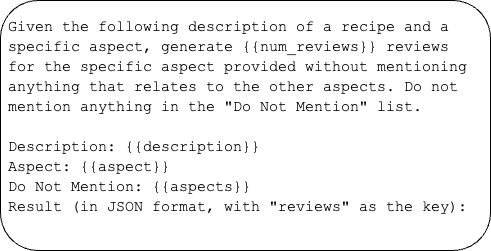}
    \caption{Disjoint Review Generation Prompt Used with GPT-4}
    \label{fig:disjoint_review_prompt}
\end{figure}

\begin{figure}[h]
    \centering
    \includegraphics[width=0.8\linewidth]{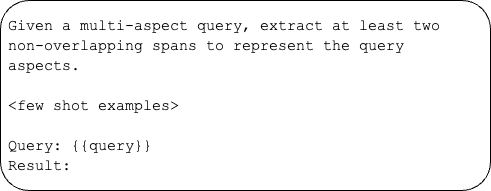}
    \caption{Query Aspect Extraction Prompt Used with GPT-4}
    \label{fig:aspect_extract_prompt}
\end{figure}

\begin{figure}[b]
    \centering
    \includegraphics[width=0.8\linewidth]{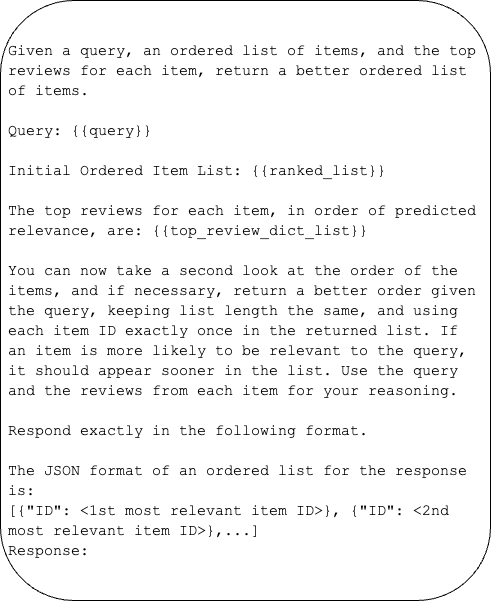}
    \caption{Generic Listwise Reranking Prompt Used with GPT-3.5}
    \label{fig:lw_rerank_prompt}
\end{figure}

We provide the prompts userd for overlapping review generation, disjoint review generation, query aspect extraction, and listwise reranking in Figures \ref{fig:overlapping_review_prompt}, \ref{fig:disjoint_review_prompt}, \ref{fig:aspect_extract_prompt}, and \ref{fig:lw_rerank_prompt} respectively.

\subsection{Results for $K_I = 5$}

In the main body we showed various results of experiments where $K_I$ was set to 10. We found that varying $K_I$ within this order of magnitude had a very small effect on the results, and therefore did not include findings for any other settings of $K_I$ above. For completeness, in this section we duplicate the preceding tables but use $K_I = 5$ instead of $K_I = 10$. See Tables \ref{tab:rq1_5},  \ref{tab:rq3_5}, \ref{tab:rq4_5}, and \ref{tab:rq2_5} for these results.

\begin{table*}
\caption{Stage 1 retriever performance for various aggregation functions and settings of $K_R$, with $K_I = 5$. All methods except Mono LF include Aspect Fusion. The values in parentheses indicate the 95\% error margin.}
\label{tab:rq1_5}
\footnotesize
\centering{}
\setlength\tabcolsep{1.5pt}
\begin{tabular}{|l|l|cc|cc|cc|cc|cc|cc|}
\hline
 & $K_R$ & \multicolumn{2}{|c|}{1} & \multicolumn{2}{|c|}{2} & \multicolumn{2}{|c|}{5} & \multicolumn{2}{|c|}{10} & \multicolumn{2}{|c|}{15} & \multicolumn{2}{|c|}{30} \\
\hline
 Dataset &   & MAP@5 & Re@5 & MAP@5 & Re@5 & MAP@5 & Re@5 & MAP@5 & Re@5 & MAP@5 & Re@5 & MAP@5 & Re@5 \\
\hline
\multirow{7}{*}{\shortstack[l]{Fully\\disjoint}} & AMean & .55 (.04) & .71 (.04) & .55 (.04) & .70 (.04) & .55 (.04) & .70 (.04) & .55 (.04) & .69 (.04) & .50 (.04) & .67 (.05) & .16 (.03) & .21 (.04) \\
 & Borda & .37 (.04) & .49 (.05) & .38 (.04) & .49 (.05) & .37 (.04) & .48 (.05) & .36 (.04) & .45 (.05) & .34 (.04) & .44 (.05) & .13 (.03) & .18 (.04) \\
 & GMean & .56 (.04) & .71 (.04) & .55 (.04) & .71 (.04) & .55 (.04) & .71 (.04) & .55 (.04) & .69 (.04) & .50 (.04) & .67 (.05) & .16 (.03) & .21 (.04) \\
 & HMean & .56 (.04) & .71 (.04) & .56 (.04) & .71 (.04) & .56 (.04) & .71 (.04) & .55 (.04) & .69 (.04) & .51 (.04) & .68 (.05) & .16 (.03) & .21 (.04) \\
 & Min & .42 (.04) & .54 (.05) & .01 (.01) & .01 (.01) & .01 (.01) & .01 (.01) & .01 (.01) & .01 (.01) & .01 (.01) & .01 (.01) & .01 (.01) & .01 (.01) \\
 & Mono LF & .39 (.04) & .56 (.05) & .39 (.04) & .57 (.05) & .40 (.04) & .57 (.05) & .40 (.04) & .58 (.05) & .42 (.04) & .61 (.05) & .39 (.04) & .58 (.05) \\
 & R-R & .25 (.03) & .53 (.05) & .25 (.03) & .53 (.05) & .25 (.03) & .54 (.05) & .26 (.03) & .54 (.05) & .22 (.03) & .47 (.05) & .07 (.02) & .15 (.03) \\
\cline{1-14}
\multirow{7}{*}{\shortstack[l]{Fully\\overlapping}} & AMean & .51 (.04) & .67 (.05) & .51 (.04) & .68 (.05) & .51 (.04) & .67 (.05) & .50 (.04) & .66 (.05) & .51 (.04) & .66 (.05) & .52 (.04) & .66 (.05) \\
 & Borda & .32 (.04) & .45 (.05) & .32 (.04) & .45 (.05) & .33 (.04) & .44 (.05) & .33 (.04) & .44 (.05) & .32 (.04) & .43 (.05) & .32 (.04) & .44 (.05) \\
 & GMean & .51 (.04) & .67 (.05) & .51 (.04) & .68 (.05) & .51 (.04) & .66 (.05) & .50 (.04) & .65 (.05) & .51 (.04) & .66 (.05) & .51 (.04) & .66 (.05) \\
 & HMean & .51 (.04) & .66 (.05) & .51 (.04) & .68 (.05) & .51 (.04) & .66 (.05) & .51 (.04) & .66 (.05) & .52 (.04) & .66 (.05) & .51 (.04) & .66 (.05) \\
 & Min & .31 (.04) & .44 (.05) & .01 (.01) & .01 (.01) & .01 (.01) & .01 (.01) & .01 (.01) & .01 (.01) & .01 (.01) & .01 (.01) & .01 (.01) & .01 (.01) \\
 & Mono LF & .49 (.04) & .66 (.05) & .50 (.04) & .66 (.05) & .50 (.04) & .66 (.05) & .50 (.04) & .67 (.05) & .49 (.04) & .65 (.05) & .48 (.04) & .65 (.05) \\
 & R-R & .20 (.03) & .50 (.05) & .22 (.03) & .50 (.05) & .22 (.03) & .50 (.05) & .21 (.03) & .50 (.05) & .21 (.03) & .48 (.05) & .22 (.03) & .49 (.05) \\
\cline{1-14}
\end{tabular}
\end{table*}

\begin{table*}
\caption{Stage 1 retriever performance by review aspect frequency and settings of $K_R$, with $K_i = 5$. ``Balanced'' refers to the fully disjoint dataset where all item aspects have the same number of reviews. The values in parentheses indicate the 95\% error margin.}
\label{tab:rq3_5}
\footnotesize
\centering{}
\setlength\tabcolsep{1.5pt}
\begin{tabular}{|l|l|cc|cc|cc|cc|cc|cc|}
\hline
 & $K_R$ & \multicolumn{2}{|c|}{1} & \multicolumn{2}{|c|}{2} & \multicolumn{2}{|c|}{5} & \multicolumn{2}{|c|}{10} & \multicolumn{2}{|c|}{15} & \multicolumn{2}{|c|}{30} \\
 Dataset &   & MAP@5 & Re@5 & MAP@5 & Re@5 & MAP@5 & Re@5 & MAP@5 & Re@5 & MAP@5 & Re@5 & MAP@5 & Re@5 \\
\hline
\multirow{2}{*}{\shortstack[l]{Balanced\\Frequency}} & AMean & .55 (.04) & .71 (.04) & .55 (.04) & .70 (.04) & .55 (.04) & .70 (.04) & .55 (.04) & .69 (.04) & .50 (.04) & .67 (.05) & .16 (.03) & .21 (.04) \\
 & Mono LF & .39 (.04) & .56 (.05) & .39 (.04) & .57 (.05) & .40 (.04) & .57 (.05) & .40 (.04) & .58 (.05) & .42 (.04) & .61 (.05) & .39 (.04) & .58 (.05) \\
\cline{1-14}
\multirow{2}{*}{\shortstack[l]{One Popular\\Aspect}} & AMean & .51 (.04) & .65 (.05) & .42 (.04) & .58 (.05) & .31 (.04) & .45 (.05) & .25 (.04) & .39 (.05) & .01 (.01) & .02 (.01) & .01 (.01) & .02 (.01) \\
 & Mono LF & .35 (.04) & .50 (.05) & .30 (.04) & .46 (.05) & .26 (.04) & .40 (.05) & .24 (.04) & .38 (.05) & .25 (.04) & .38 (.05) & .25 (.04) & .38 (.05) \\
\cline{1-14}
\multirow{2}{*}{\shortstack[l]{One Rare\\Aspect}} & AMean & .51 (.04) & .65 (.05) & .43 (.04) & .59 (.05) & .35 (.04) & .48 (.05) & .33 (.04) & .45 (.05) & .15 (.03) & .20 (.04) & .06 (.02) & .08 (.03) \\
 & Mono LF & .37 (.04) & .53 (.05) & .34 (.04) & .51 (.05) & .30 (.04) & .45 (.05) & .28 (.04) & .43 (.05) & .30 (.04) & .45 (.05) & .28 (.04) & .40 (.05) \\
\cline{1-14}
\end{tabular}
\end{table*}

\begin{table*}
\caption{Stage 1 retriever performance by whether labelled GT or extracted query aspects are used, with $K_I = 5$. The values in parentheses indicate the 95\% error margin.}
\label{tab:rq4_5}
\footnotesize
\centering{}
\setlength\tabcolsep{1.5pt}
\begin{tabular}{|l|l|cc|cc|cc|cc|cc|cc|}
\hline
 & $K_R$ & \multicolumn{2}{|c|}{1} & \multicolumn{2}{|c|}{2} & \multicolumn{2}{|c|}{5} & \multicolumn{2}{|c|}{10} & \multicolumn{2}{|c|}{15} & \multicolumn{2}{|c|}{30} \\
 &   & MAP@5 & Re@5 & MAP@5 & Re@5 & MAP@5 & Re@5 & MAP@5 & Re@5 & MAP@5 & Re@5 & MAP@5 & Re@5 \\
\hline
\multirow{2}{*}{\shortstack[l]{Extracted\\query aspects}} & AMean & .45 (.04) & .60 (.05) & .46 (.04) & .63 (.05) & .46 (.04) & .62 (.05) & .45 (.04) & .61 (.05) & .44 (.04) & .62 (.05) & .14 (.03) & .20 (.04) \\
 & Mono LF & .39 (.04) & .56 (.05) & .39 (.04) & .57 (.05) & .40 (.04) & .57 (.05) & .40 (.04) & .58 (.05) & .42 (.04) & .61 (.05) & .39 (.04) & .58 (.05) \\
\cline{1-14}
\multirow{2}{*}{\shortstack[l]{GT query\\aspects}} & AMean & .55 (.04) & .71 (.04) & .55 (.04) & .70 (.04) & .55 (.04) & .70 (.04) & .55 (.04) & .69 (.04) & .50 (.04) & .67 (.05) & .16 (.03) & .21 (.04) \\
 & Mono LF & .39 (.04) & .56 (.05) & .39 (.04) & .57 (.05) & .40 (.04) & .57 (.05) & .40 (.04) & .58 (.05) & .42 (.04) & .61 (.05) & .39 (.04) & .58 (.05) \\
\cline{1-14}
\end{tabular}
\end{table*}

\begin{table*}
\caption{Stage 2 reranker performance by reranking method and setting of $K_R$, with $K_I = 5$. ``No'' refers to the case where no reranking is applied, and is equivalent to the stage 1 results. The values in parentheses indicate the 95\% error margin.}
\label{tab:rq2_5}
\footnotesize
\centering{}
\setlength\tabcolsep{1.5pt}
\begin{tabular}{|l|l|l|cc|cc|cc|cc|cc|cc|}
\hline
 &  & $K_R$ & \multicolumn{2}{|c|}{1} & \multicolumn{2}{|c|}{2} & \multicolumn{2}{|c|}{5} & \multicolumn{2}{|c|}{10} & \multicolumn{2}{|c|}{15} & \multicolumn{2}{|c|}{30} \\
 &  &   & MAP@5 & Re@5 & MAP@5 & Re@5 & MAP@5 & Re@5 & MAP@5 & Re@5 & MAP@5 & Re@5 & MAP@5 & Re@5 \\
\hline
\multirow{6}{*}{\shortstack[l]{Fully\\disjoint}} & \multirow{3}{*}{AMean} & CE & .41 (.04) & .71 (.04) & .52 (.04) & .70 (.04) & .53 (.04) & .70 (.04) & .53 (.04) & .69 (.04) & .53 (.04) & .67 (.05) & .17 (.03) & .21 (.04) \\
 &  & LW & .44 (.04) & .71 (.04) & .48 (.04) & .70 (.04) & .53 (.04) & .70 (.04) & .55 (.04) & .69 (.04) & .53 (.04) & .67 (.05) & .16 (.03) & .21 (.04) \\
 &  & No & .55 (.04) & .71 (.04) & .55 (.04) & .70 (.04) & .55 (.04) & .70 (.04) & .55 (.04) & .69 (.04) & .50 (.04) & .67 (.05) & .16 (.03) & .21 (.04) \\
\cline{2-15}
 & \multirow{3}{*}{Mono LF} & CE & .35 (.04) & .56 (.05) & .37 (.04) & .57 (.05) & .38 (.04) & .57 (.05) & .41 (.04) & .58 (.05) & .46 (.04) & .61 (.05) & .46 (.04) & .58 (.05) \\
 &  & LW & .33 (.04) & .56 (.05) & .34 (.04) & .57 (.05) & .37 (.04) & .57 (.05) & .37 (.04) & .58 (.05) & .44 (.04) & .61 (.05) & .43 (.04) & .58 (.05) \\
 &  & No & .39 (.04) & .56 (.05) & .39 (.04) & .57 (.05) & .40 (.04) & .57 (.05) & .40 (.04) & .58 (.05) & .42 (.04) & .61 (.05) & .39 (.04) & .58 (.05) \\
\cline{1-15} \cline{2-15}
\multirow{6}{*}{\shortstack[l]{Fully\\overlapping}} & \multirow{3}{*}{AMean} & CE & .51 (.04) & .67 (.05) & .51 (.04) & .68 (.05) & .51 (.04) & .67 (.05) & .50 (.04) & .66 (.05) & .49 (.04) & .66 (.05) & .51 (.04) & .66 (.05) \\
 &  & LW & .48 (.04) & .67 (.05) & .50 (.04) & .68 (.05) & .52 (.04) & .67 (.05) & .51 (.04) & .66 (.05) & .53 (.04) & .66 (.05) & .53 (.04) & .66 (.05) \\
 &  & No & .51 (.04) & .67 (.05) & .51 (.04) & .68 (.05) & .51 (.04) & .67 (.05) & .50 (.04) & .66 (.05) & .51 (.04) & .66 (.05) & .52 (.04) & .66 (.05) \\
\cline{2-15}
 & \multirow{3}{*}{Mono LF} & CE & .50 (.04) & .66 (.05) & .50 (.04) & .66 (.05) & .50 (.04) & .66 (.05) & .50 (.04) & .67 (.05) & .48 (.04) & .65 (.05) & .49 (.04) & .65 (.05) \\
 &  & LW & .47 (.04) & .66 (.05) & .48 (.04) & .66 (.05) & .52 (.04) & .66 (.05) & .53 (.04) & .67 (.05) & .51 (.04) & .65 (.05) & .51 (.04) & .65 (.05) \\
 &  & No & .49 (.04) & .66 (.05) & .50 (.04) & .66 (.05) & .50 (.04) & .66 (.05) & .50 (.04) & .67 (.05) & .49 (.04) & .65 (.05) & .48 (.04) & .65 (.05) \\
\cline{1-15} \cline{2-15}
\end{tabular}
\end{table*}

\subsection{Data for Figure \ref{fig:rq2_bubble}}

In Figure \ref{fig:rq2_bubble}, we show the number of queries for which the correct item was ranked in a certain position by the stage 1 retriever and stage 2 reranker. The underlying data for this figure is shown in Table \ref{tab:bubble_data}.

\begin{table*}[t]
\caption{Ranks assigned to the correct items for stage 1 retriever and stage 2 CE reranker with AMean aggregation Aspect Fusion.}
\label{tab:bubble_data}
\footnotesize
\centering{}
\begin{tabular}{|c|r|cccccccccc|}
\hline
 & & \multicolumn{10}{c|}{Stage 2 Correct Item Rank} \\
\hline
  & \shortstack[l]{Stage 1 Correct\\Item Rank} & 1 & 2 & 3 & 4 & 5 & 6 & 7 & 8 & 9 & 10 \\
\hline
\multirow{10}{*}{$K_R = 1$} & 1 & 68 & 19 & 12 & 9 & 2 & 2 & 4 & 2 & 0 & 0 \\
 & 2 & 12 & 20 & 8 & 1 & 3 & 4 & 1 & 3 & 0 & 0 \\
 & 3 & 5 & 3 & 8 & 5 & 2 & 2 & 2 & 1 & 1 & 1 \\
 & 4 & 1 & 2 & 2 & 4 & 1 & 4 & 2 & 0 & 1 & 0 \\
 & 5 & 1 & 1 & 1 & 2 & 3 & 0 & 2 & 2 & 1 & 1 \\
 & 6 & 0 & 0 & 0 & 1 & 0 & 0 & 1 & 1 & 1 & 1 \\
 & 7 & 3 & 3 & 2 & 2 & 1 & 3 & 1 & 1 & 3 & 0 \\
 & 8 & 1 & 0 & 2 & 0 & 1 & 0 & 1 & 0 & 4 & 0 \\
 & 9 & 0 & 0 & 0 & 0 & 1 & 0 & 1 & 0 & 2 & 0 \\
 & 10 & 0 & 1 & 0 & 1 & 1 & 1 & 2 & 1 & 0 & 1 \\
\cline{1-12}
\multirow{10}{*}{$K_R = 30$} & 1 & 83 & 21 & 5 & 5 & 0 & 2 & 1 & 0 & 0 & 0 \\
 & 2 & 29 & 12 & 2 & 1 & 2 & 2 & 3 & 1 & 1 & 0 \\
 & 3 & 12 & 5 & 3 & 3 & 2 & 0 & 1 & 2 & 0 & 0 \\
 & 4 & 7 & 2 & 1 & 1 & 1 & 3 & 1 & 0 & 0 & 0 \\
 & 5 & 7 & 2 & 7 & 5 & 0 & 0 & 1 & 1 & 0 & 0 \\
 & 6 & 5 & 1 & 3 & 1 & 0 & 2 & 2 & 0 & 0 & 0 \\
 & 7 & 3 & 1 & 1 & 1 & 0 & 1 & 0 & 0 & 0 & 1 \\
 & 8 & 4 & 0 & 0 & 1 & 0 & 2 & 1 & 0 & 0 & 0 \\
 & 9 & 0 & 2 & 1 & 0 & 1 & 1 & 0 & 1 & 0 & 0 \\
 & 10 & 1 & 2 & 1 & 0 & 0 & 0 & 1 & 1 & 0 & 0 \\
\cline{1-12}
\end{tabular}
\end{table*}

\end{document}